# Superconductivity at 6 K and the violation of Pauli limit in Ta$_2$Pd$_x$S$_5$

Y. F. Lu[1], T. Takayama[2], A. F. Bangura[3], Y. Katsura[4], D. Hashizume[3] and H. Takagi[2,4]

[1]Department of Advanced Materials, University of Tokyo, Kashiwa, Chiba 277-8561, Japan

[2] Max Planck Institute for Solid State Research, Heisenbergstrasse 1, 70569 Stuttgart, Germany

[3]RIKEN Advanced Science Institute, Wako, Saitama 351-0198, Japan

[4]Department of Physics, University of Tokyo, Hongo, Tokyo 113-0033, Japan

Ta$_2$Pd$_x$S$_5$ ($x \lesssim 1.0$) was found to show superconductivity at $T_c \sim 6$ K. The temperature dependent resistivity of single crystalline Ta$_2$Pd$_{0.92}$S$_5$ showed that the system is strongly disordered due to Pd deficiencies and close to Anderson localized state. Superconductivity in the dirty limit as well as the temperature dependence of specific heat $C(T)$ implies that superconductivity is $s$-wave. The upper critical field $\mu_0 H_{c2}$ at $T = 0$ K limit with the magnetic field parallel to the TaS$_6$ chains was found to be as high as 31 T, exceeding the Pauli paramagnetic limit $\mu_0 H_\mathrm{P} = 10.2$ T by a factor of 3. We argue that the absence of the paramagnetic pair-breaking originates from strong spin-orbit scattering due to Pd deficiencies embedded in the periodic lattice of heavy 5$d$ Ta and 4$d$ Pd.

Superconducting 4$d$ and 5$d$ compounds recently gained a renewed interest because of the strong spin-orbit coupling inherent to heavy elements. In non-centrosymmetric 4$d$ and 5$d$ superconductors, pronounced admixture of different parity in the order parameter is expected and, indeed, a gapless superconductivity due to such admixture has been discussed in Li$_2$Pt$_3$B [1] and CePt$_3$Si [2] which both have breaking inversion symmetry in their crystal structure. Even in the presence of the inversion symmetry, spin-orbit coupling may give rise to an exotic superconductivity, for example associated with the local breaking of inversion symmetry and, as we show in this letter, through strong spin-orbit scattering.

We discovered a new superconductor Ta$_2$Pd$_x$S$_5$ ($x \lesssim 1.0$) with $T_c \sim 6$ K. Ta$_2$Pd$_x$S$_5$ was first synthesized by Squattorito et al. and reported to be Pd deficient Ta$_2$Pd$_{0.89}$S$_5$ [3]. The crystal structure is illustrated in the inset of Fig. 1 [4]. Ta atoms are prismatically coordinated by six sulfur atoms, and the TaS$_6$ prisms form one-dimensional chains along the $b$-axis by sharing their edges. Pd atoms bridge the TaS$_6$ chains by forming the PdS$_4$ squares in between. There are two Pd sites, and one of them, labeled as Pd(2), is not fully occupied.

In this letter, we demonstrate that Ta$_2$Pd$_x$S$_5$ is a dirty limit $s$-wave superconductor. Despite seemingly conventional superconductivity, the sulfide with heavy transition metals shows an extremely large upper critical field $\mu_0 H_{c2}(0) \sim 31$ T and violates the Pauli paramagnetic limit by a factor of 3. We argue that strong spin-orbit scattering is caused by the Pd defects on the periodic lattice of heavy 5$d$ Ta and 4$d$ Pd, which suppresses the paramagnetic pair breaking almost completely.

The polycrystalline samples used in this study were synthesized by a conventional solid state reaction. Powders of elemental Ta, Pd and S were mixed and pelletized with a starting composition of 2 : $x_{nom}$ : 5, where $x_{nom}$ denotes nominal Pd content varying from 0.99 to 1.10 [5]. The pellets were sealed in an Ar-filled quartz tube, and slowly heated up to 400 °C to avoid rapid volatilization of sulfur. They were subsequently heated to 820 °C and calcined for 30 hours. The obtained samples were air-stable. The x-ray diffraction pattern of the $Ta_2Pd_xS_5$ polycrystalline sample showed that the obtained powder was almost single phase with a small trace of $TaS_2$, as shown in Fig. 1. The result of Rietveld refinement indicated the presence of Pd deficiency in $Ta_2Pd_xS_5$ and hereafter we use Pd content $x$ determined from the Rietveld refinement to represent the composition of the samples [6]. A systematic change of lattice constants as a function of $x$ was observed, indicating that the amount of Pd deficiencies can be controlled.

Single crystals of $Ta_2Pd_xS_5$ were obtained by a chemical vapor transport using $I_2$ as a transporting agent. Black needle-shaped crystals were grown and elongated along the *b*-axis. The structure of the obtained crystals was refined by an x-ray diffractometer using Mo-$K\alpha$ radiation, as summarized in Table 1. The amount of Pd in the single crystal was determined to be 0.92, which supports the presence of Pd deficiencies [7].

Electric, magnetic, and thermodynamic properties of the polycrystalline and single crystalline samples were characterized by using Quantum design PPMS and MPMS. High field measurements up to 35 T were conducted on the single crystals at NHMFL Florida.

Electrical resistivity and magnetization measurements showed that $Ta_2Pd_xS_5$ is a paramagnetic metal. Provided that $Ta_2PdS_5$ were totally ionic, the valence states could be $Ta^{5+}$, $Pd^0$, and $S^{2-}$, where $Ta^{5+}$ and $Pd^{0+}$ have empty $5d^0$ and fully occupied $4d^{10}$ configurations, respectively. The electronic structure calculation using WIEN2k [8] showed that Ta $5d$ and Pd $4d$ (S$3p$) bands in fact overlap and the system is a compensated metal with a charge transfer from Pd(S) to Ta [9].

$Ta_2Pd_xS_5$ was found to exhibit bulk superconductivity at $T_c \sim 6$ K as evidenced by zero resistivity and Meissner signal shown in Figs. 2(a), (b). $T_c$ gradually decreases with increasing Pd deficiency.

The electronic and phononic contributions to the specific heat were estimated with the standard model, $C(T) = \gamma T + \beta T^3$ for the sample with $x = 0.97$, by fitting the $C(T)$ under 12 T shown in the inset of Fig. 2(c). This yielded the coefficients of electronic specific heat $\gamma = 27.6$ mJ/mol·K$^2$ and of the lattice term $\beta = 1.25$ mJ/mol·K$^4$. The obtained $\gamma$ is relatively large for a metal, which suggests that $Ta_2Pd_xS_5$ hosts moderately heavy mass carriers. The Debye temperature $\Theta_D$ was calculated to be 231 K. The electronic specific heat $C_e(T)$ estimated by subtracting the lattice term is depicted in Fig. 2(c). A jump in $C_e(T)$ was seen around $T_c \sim 5.8$ K and was broadened probably due to a chemical inhomogeneity hindering our ability to do a reasonable fit of $C_e(T)$ with BCS theory. The rapidly diminishing $C_e(T)$ well below $T_c$, however, strongly suggests gapful superconductivity. Combining this with the presence of strong disorder described below which would suppress any sign-changing order parameter, we conclude that $s$-wave superconductivity is realized in $Ta_2Pd_xS_5$.

The resistivity $\rho(T)$ of single crystals of $Ta_2Pd_{0.92}S_5$, measured along the chains (// the $b$-axis) show a superconducting transition at 5.4 K, consistent with the polycrystalline data. The magnitude of $\rho(T)$ right above $T_c$ was as high as 0.3 mΩcm, despite that the measurement was conducted along very likely the most conductive direction. A small upturn in $\rho(T)$ at low temperatures can be seen in the inset of Fig. 3, implying that the system is close to an Anderson insulator associated with the strong disorder due to the Pd defects. Let us assume for simplicity 1 electron in each of the two Ta bands (total 2 electrons) and 2 holes in the Pd band due to the band overlap and that they have the same effective mass with free-electron like dispersion. The transport mean free path $l_{tr}$ was roughly estimated to be 0.4 nm from the magnitude of $\rho(T)$ at low temperature. The estimate depends on the assumption of carrier density and the effective masses but, within reasonable parameter space, $l_{tr}$ should be of the order of 1 nm. The extremely short mean free path, of the order of a few lattice spacing, is consistent with the weakly localized behavior of $\rho(T)$.

In a seemingly conventional $s$-wave superconductor with strong disorder, we found a substantial violation of Pauli limit in the upper critical field $\mu_0 H_{c2}(T)$ of $Ta_2Pd_xS_5$. The $\mu_0 H_{c2}(T)$ were evaluated from the magnetoresistive transitions with magnetic fields parallel and perpendicular to the $b$-axis of the single crystalline $Ta_2Pd_{0.92}S_5$ shown in Fig. 3. The anisotropy in the upper critical fields can be clearly seen; the upper critical field with a field applied along the $b$-axis ($TaS_6$ chains) is larger than that with a field applied perpendicular to the *b*-axis, likely reflecting the anisotropic Fermi velocities. At high fields above $\mu_0 H_{c2}(T)$, the normal state resistivity showed negative magnetoresistance that becomes more pronounced at lower temperature and is probably associated with a weak localization. This is in accord with the negative temperature coefficient of $\rho(T)$ observed right above $T_c$.

The $\mu_0 H_{c2}(T)$ obtained from the midpoints of magnetoresistive transition is plotted in the inset of Fig. 4. The slopes $\mu_0 \frac{dH_{c2}}{dT}\bigg|_{T=T_c}$ were -8.5 T/K and -3.7 T/K for magnetic fields parallel to and perpendicular to the *b*-axis, respectively. From the slopes, the orbital limit upper critical field $\mu_0 H_{c2}^*(0)$ can be estimated based on Werthamer-Helfand-Hohenberg (WHH) theory for a single band BCS superconductor,

$$\mu_0 H_{c2}^*(0) = -0.698 \times \mu_0 \frac{dH_{c2}}{dT}\bigg|_{T=T_c} \times T_c$$ [10]. The obtained $\mu_0 H_{c2}^*(0)$ are 32 T (*H* // *b*) and 14 T (*H* ⊥ *b*), are very high for a superconductor with modest critical temperature $T_c$ = 5.4 K. The Ginzburg-Landau coherence lengths $\xi_{GL}$ ~ 3.2 nm (*H* // *b*) and $\xi_{GL}$ ~ 4.8 nm (*H* ⊥ *b*), estimated from the orbital limit $\mu_0 H_{c2}^*(0)$, are longer than the estimated $l_{tr}$, evidencing that superconductivity in Ta$_2$Pd$_{0.92}$S$_5$ is indeed in the dirty limit. Since these coherence lengths, obtained from $\mu_0 H_{c2}^*(0)$, are longer than the distance between the TaS$_6$ chains, superconductivity in this compound is anisotropic but three-dimensional in nature.

In dirty-limit BCS superconductors, the slope of upper critical field near $T_c$ is related to the normal state parameters as $\mu_0 \frac{dH_{c2}}{dT}\bigg|_{T=T_c}$ [T/K] $= 4.44 \times \gamma$ [erg/K$^2$cm$^3$] $\times \rho_0$ [$\Omega$cm]. The experimental values of $\rho_0 = 0.3$ m$\Omega$cm and $\gamma = 3127$ erg/K$^2$cm$^3$ yield $\mu_0 \frac{dH_{c2}}{dT}\bigg|_{T=T_c} =$ -4.2 T/K, close to the experimentally observed slope. This means the upper critical field slope can be understood within the frame work of dirty limit BCS superconductor and the large slope can be ascribed to the pronounced disorder (large $\rho$) and the sizable effective mass (large $\gamma$).

The large slope of $\mu_0 H_{c2}(T)$ gives us an opportunity to challenge the Pauli limit. Superconductivity may be destroyed by paramagnetic pair breaking when paramagnetic Zeeman energy $1/2\chi_P H^2$ overcomes superconducting condensation energy $1/2N(0)\Delta^2$ [11], where $\chi_P$, $N(0)$, $\Delta$ denote Pauli paramagnetic susceptibility, density of states at the Fermi level and superconducting gap, respectively. In a BCS superconductor in the weak-coupling limit, the Pauli limit is $\mu_0 H_P$ [T] $= 1.84 \times T_c$ [K], which corresponds to 10.2 T for the Ta$_2$Pd$_{0.92}$S$_5$ single crystal. The Pauli limiting behavior of $\mu_0 H_{c2}(T)$ has been observed in high field superconductors including Fe(Se, Te) [12] and $\alpha$-(ET)$_2$NH$_4$(SCN)$_4$ [13]. In a stark contrast with these compounds, the measured upper critical fields of Ta$_2$Pd$_{0.92}$S$_5$ at low temperatures surpassed the Pauli limit as shown in Fig. 4. The extrapolated $\mu_0 H_{c2}(0)$ appear to be 31 T and 14 T, for magnetic fields parallel and perpendicular to the $b$-axis, respectively. Note that $\mu_0 H_{c2}(0)$ along the $b$-axis violates the Pauli limit by a factor of 3. The experimentally observed $\mu_0 H_{c2}(0)$ are not very different to the orbital limit $\mu_0 H_{c2}^*(0) = 32$ T estimated from the WHH theory, implying that paramagnetic pair breaking effect is suppressed almost completely in Ta$_2$Pd$_x$S$_5$.

There are several factors which could suppress paramagnetic pair breaking. First, in *p*-wave superconductors with spin-triplet pairing such as (TMTSF)$_2$PF$_6$ [14] and URhGe [15], there should be no paramagnetic pair breaking. However, Ta$_2$Pd$_x$S$_5$ is an almost localized and dirty limit superconductor, where spin-triplet channel should be completely suppressed. Non-centrosymmetric superconductors with spin-orbit coupling should have admixture of spin triplet component in their order parameter, which could suppress the paramagnetic pair breaking as in *p*-wave superconductors [16]. The crystal structure of Ta$_2$Pd$_x$S$_5$ includes an inversion center (Space group *C*2/*m*, No.12), and is thus incompatible with the unconventional pairing possible in non-centrosymmetric systems. It was suggested recently, however, that local breaking of inversion symmetry might produce admixture of unconventional pairing state, despite retaining macroscopic space inversion symmetry [17, 18]. The local site symmetry of 5*d* Ta in Ta$_2$Pd$_x$S$_5$ breaks the space inversion symmetry. However, again, the exotic pairing component should be completely suppressed by the strong disorder.

Strong-coupling superconductivity where superconducting gap is enhanced compared with that of weak coupling BCS would enhance the Pauli limit [19]. The enhancement by a factor of 3 is too large to be accounted for based on the strong coupling which could enhance the gap up to a factor of 1.5.

A reduced *g*-factor may suppress the paramagnetic pair breaking, which was discussed to be the case for URu$_2$Si$_2$ [20]. This effect could be pronounced in superconductors with heavy elements such as 5$d$ transition metals. The magnetic susceptibility of the Ta$_2$Pd$_{0.97}$S$_5$ polycrystalline sample in the normal state was almost temperature independent, dominated by Pauli paramagnetism. The temperature independent magnetic susceptibility was 3.8 × 10$^{-4}$ emu/mol. By subtracting core diamagnetism which are obtained from ionic values of Ta$^{5+}$, Pd$^{2+}$ and S$^{2-}$, the paramagnetic susceptibility was estimated as $\chi_P$ ~ 6.0 × 10$^{-4}$ emu/mol . Combined with the electronic specific heat coefficient $\gamma$ of 27.6 mJ/mol·K$^2$, the Wilson ratio was evaluated to be 1.6, moderately enhanced compared with the free electron value of 1. The moderately enhanced Wilson ratio indicates that the pronounced reduction of $g$-factor is unlikely the case.

From the consideration above, we argue that the spin-orbit scattering, associated with the presence of heavy Ta and Pd atoms, plays a central role in suppressing the paramagnetic pair breaking. Spin-orbit scattering gives rise to a remnant spin paramagnetic susceptibility in the superconducting state, and therefore suppresses the paramagnetic pair breaking effect. This effect can be described within the WHH theory by a parameter $\lambda_{so} = 2\hbar/3\pi k_B T_c \tau_{so} = \xi_0/l_{so}$, where $\tau_{so}$, $l_{so}$ denote a relaxation time and mean free path of spin-orbit scattering and $\xi_0$ is BCS coherence length. With $\lambda_{so}$, the upper critical field $\mu_0 H_{c2}(T)$ is determined by the following equation,

$$\ln\left(\frac{1}{t}\right) = \left(\frac{1}{2} + \frac{i\lambda_{so}}{4\gamma}\right)\psi\left(\frac{1}{2} + \frac{h + \lambda_{so}/2 + i\gamma}{2t}\right) + \left(\frac{1}{2} - \frac{i\lambda_{so}}{4\gamma}\right)\psi\left(\frac{1}{2} + \frac{h + \lambda_{so}/2 - i\gamma}{2t}\right) - \psi\left(\frac{1}{2}\right)$$

where $\psi(x)$ is a digamma function, $t = T/T_c$, $h = 0.281 H_{c2}(T)/H_{c2}^*(0)$ and $\gamma =$

$(\alpha^2 h^2 - \lambda_{so}^2/4)^{1/2}$ [10]. $\alpha = \sqrt{2}H_{c2}^*(0)/H_P$, a ratio of the orbital limiting critical field and the Pauli limiting field called Maki parameter, gives a measure of paramagnetic effect. In $Ta_2Pd_{0.92}S_5$, $\alpha$ was estimated as 4.4 ($\mu_0 H_{c2}^*(0)$ = 32 T and $\mu_0 H_P$ = 10.2 T). By tuning $\lambda_{so}$ as a fitting parameter, we obtain a large $\lambda_{so}$ ~ 100 to reproduce $\mu_0 H_{c2}(T)$. The large $\lambda_{so}$ simply reflects that $\mu_0 H_{c2}(T)$ behaves as if there were no Pauli paramagnetic pair breaking. We note that the estimate of $\lambda_{so}$ here should be taken as an order estimate because the effects of multiband and inhomogeneity were neglected [21] and the fitting is relatively insensitive to $\lambda_{so}$ when $\lambda_{so} \gg 1$.

$\lambda_{so}$ of the order of ~100 means extremely strong spin-orbit scattering and short $l_{so}$. In dirty limit superconductors, $\xi_0$ is related with $\xi_{GL}$ as $\xi_{GL}^2 \sim (0.85)^2 l_{tr} \cdot \xi_0$. With $\xi_{GL}$ ~ 3.2 nm for $\mu_0 H_{c2}^*(0)$ = 32 T observed in $Ta_2Pd_{0.92}S_5$, $\lambda_{so}$ ~100 yields $l_{tr} \cdot l_{so} \sim \xi_{GL}^2/8.5^2$ ~ $(0.4 \text{ nm})^2$. Given a rough estimate of $l_{tr}$ ~ 0.4 nm, $l_{so}$ should be of the order of ~ 1 nm and hence comparable to $l_{tr}$. This implies that spin-orbit scattering is as strong as transport scattering by Pd defects.

The effect of spin-orbit scattering is profound when heavy mass impurities with strong spin-orbit coupling are present. The suppression of paramagnetic pair breaking due to heavy mass impurities were discussed in variety of superconductor partly motivated by the possibility of increasing the upper critical field [22]. In the case of $Ta_2Pd_xS_5$, Pd deficiencies, the main cause of transport scattering, are not heavy mass impurities. Very strong spin-orbit coupling is expected for 5$d$ Ta and 4$d$ Pd. They, however, form a periodic lattice and do not contribute as spin-orbit scatterers. We argue that Pd deficiencies (i.e. the atomic number $Z = 0$) elements embedded in the lattice of heavy elements may act analogous to heavy elements in a lattice of light elements. Pd deficiencies in this case scatter electrons and holes in $5d$ and $4d$ bands where the spins are coherently and strongly entangled with the momentum. They could be strong spin-orbit scatterers as well as transport scatterers. In the dirty limit, such spin-orbit scattering should be maximized.

Within the above scenario, the strong spin-orbit scattering and the resultant suppression of Pauli limiting should be generic to strongly disordered $4d$ and $5d$ based high field superconductors. In fact, the violation of Pauli limit was observed in dirty high field superconductors like Chevrel phase $PbMo_6S_8$ and $SnMo_6S_8$ [23]. It might be interesting to examine the recently observed violation of Pauli limit in $Li_{0.9}Mo_6O_{17}$ from the spin-orbit scattering limit where negative temperature dependence of $\rho(T)$ was observed [24].

In summary, we discovered a new superconductor $Ta_2Pd_xS_5$ with $T_c \sim 6$ K. The short mean free path $l_{tr}$ and moderately large mass lead to a large upper critical field $\mu_0H_{c2}(0)$ ~31 T along the $b$-axis. This $\mu_0H_{c2}(0)$ violates the Pauli limit by a factor of 3, which is likely associated with strong spin-orbit scattering induced by Pd defects acting as a "spin-orbit vacancy". This suggests that $5d$ and $4d$ based superconductors can provide an opportunity to substantialize new high-field superconductors when the orbital-limit can be enhanced.

We thank L. Balicas and G. Li for technical support, and G. Khaliullin, G. Jackeli, I. Mazin and A. Schnyder for invaluable discussions. This work was partly supported by Grant-in-Aid for Scientific Research (S) (Grant No. 24224010), and the work at NHMFL is supported by NSF and state of Florida.

247004 (2005).

[17] Y. Nishikubo, K. Kudo, M. Nohara, J. Phys. Soc. Jpn. **80**, 055002 (2011).

[18] S. J. Youn, M. H. Fischer, S. H. Rhim, M. Sigrist, and D. F. Agterberg, Phys. Rev. B **85**, 220505(R) (2012).

[19] T. P. Orlando, E. J. McNiff Jr, S. Foner and M. R. Beasley, Phys. Rev. B, **19**, 4545 (1979).

[20] M. M. Altarawneh, N. Harrison, G. Li, L. Balicas, P. H. Tobasha, F. Ronning and E. D. Bauer, Phys. Rev. Lett., 108, 066407 (2012).

[21] L. Coffey, K. A. Muttalib, and K. Levin, Phys. Rev. Lett. **52**, 783 (1984).

[22] L. J. Neuringer and Y. Shapira, Phys. Rev. Lett., **17**, 81 (1966).

[23] Ø. Fischer, Appl. Phys. **16**, 1 (1978).

[24] J. –F. Mercure, A. F. Bangura, Xiaofeng Xu, N. Wakeham, A. Carrington, P. Walmsley, M. Greenblatt, and N. E. Hussey, Phys. Rev. Lett. **108**, 187003 (2012).
Table 1. Refined structural parameters of single crystalline $Ta_2Pd_{0.92}S_5$. The space group is $C2/m$ (No. 12) and $Z = 4$, and the lattice parameters are $a$ = 12.265(3) Å, $b$ = 3.2793(6) Å, $c$ = 15.262(3) Å, $\alpha = \gamma = 90°$, and $\beta = 104.883(3)°$. $g$ and $U_{iso}$ denote site occupancy and equivalent isotropic displacement parameters, respectively. The final $R$ indices after the refinements are $R(F) = 0.0187$ and $wR(F^2) = 0.0411$.

Fig. 1 X-ray diffraction pattern of the Ta$_2$Pd$_{0.97}$S$_5$ polycrystalline sample recorded by using Cu-$K\alpha$ radiation at room temperature. The red crosses, black solid line, lower green bars and blue solid line represent the experimental data, calculated pattern, expected peak positions and the difference between the experimental data and calculated pattern, respectively. The inset depicts the crystal structure of Ta$_2$Pd$_x$S$_5$. Ta, Pd(1), Pd(2) and S atoms are represented as red, blue, green and yellow spheres. The refined lattice parameters are $a$ = 12.273(3) Å, $b$ = 3.280(6) Å, $c$ = 15.294(8) Å, $\alpha = \gamma =$ 90°, and $\beta$ = 105.20(2)°.

Fig. 2 Superconductivity observed in Ta$_2$Pd$_x$S$_5$ polycrystalline samples with different $x$. (a) Resistivity. (b) Magnetization. (c) Electronic specific heat divided by temperature. ZFC and FC in (b) represent zero-field cooling and field-cooling sequences, respectively. The inset in (c) shows specific heats under magnetic fields of 0 and 12 T. The broken line indicates the fitting line for the data under 12 T based on the standard model $C(T)/T = \gamma + \beta T^2$.

Fig. 3 Resistive superconducting transition of Ta$_2$Pd$_{0.92}$S$_5$ single crystal ($T_c$ = 5.4 K) with applying magnetic fields (a) parallel or (b) perpendicular to the $b$-axis, respectively. In (a), the data were collected at $T$ = 0.50, 1.32, 1.53, 2.07, 2.43, 2.77, 3.24, 3.58, 3.78 and 4.50 K, whereas at $T$ = 0.55, 1.27, 1.33, 1.81, 2.10, 2.78, 3.17, 3.62 and 4.30 K in (b). The inset shows the temperature dependence of resistivity of Ta$_2$Pd$_{0.92}$S$_5$ single crystal along the $b$-axis.

Fig. 4 Temperature dependence of upper critical field $\mu_0H_{c2}(T)$ with field parallel to the

$b$-axis. The dashed lines represent the fitting lines based on the WHH theory with Maki parameter $\alpha = 4.4$, and spin-orbit scattering rate $\lambda_{so} = 0$, 2, 10 and 100. The inset shows $\mu_0 H_{c2}(T)$ with magnetic fields applied parallel or perpendicular to the $b$-axis. The black broken line in the inset represents the nominal Pauli limit $\mu_0 H_P = 10.2$ T.

Table 1

| Atom | site | g | x | y | z | $U_{iso}$ (Å$^2$) |
|---|---|---|---|---|---|---|
| Ta(1) | 4$i$ | 1 | 0.0744(1) | 1/2 | 0.1824(1) | 0.005(1) |
| Ta(2) | 4$i$ | 1 | 0.1506(1) | 0 | 0.3767(1) | 0.004(1) |
| Pd(1) | 2$a$ | 1 | 0 | 0 | 0 | 0.007(1) |
| Pd(2) | 2$c$ | 0.838(2) | 0 | 0 | 1/2 | 0.008(1) |
| S(1) | 4$i$ | 1 | 0.3510(1) | 0 | 0.4944(1) | 0.006(1) |
| S(2) | 4$i$ | 1 | 0.2508(1) | 1/2 | 0.3045(1) | 0.005(1) |
| S(3) | 4$i$ | 0.985(5) | 0.1731(1) | 0 | 0.1096(1) | 0.006(1) |
| S(4) | 4$i$ | 1 | 0.4236(1) | 1/2 | 0.1249(1) | 0.007(1) |
| S(5) | 4$i$ | 1 | 0.4984(1) | 0 | 0.3158(1) | 0.006(1) |

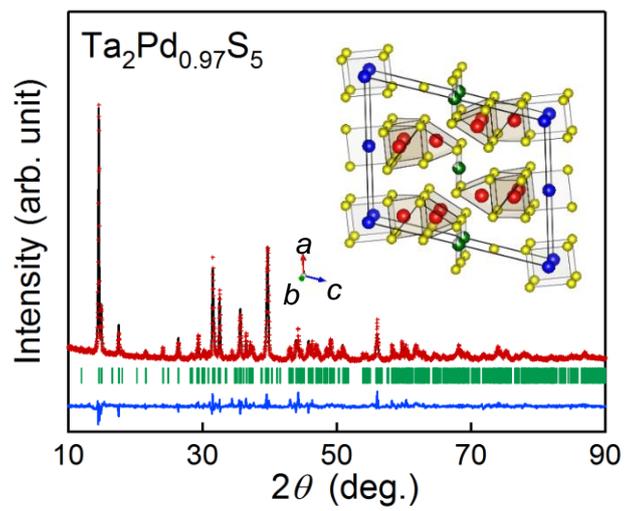

Fig. 1

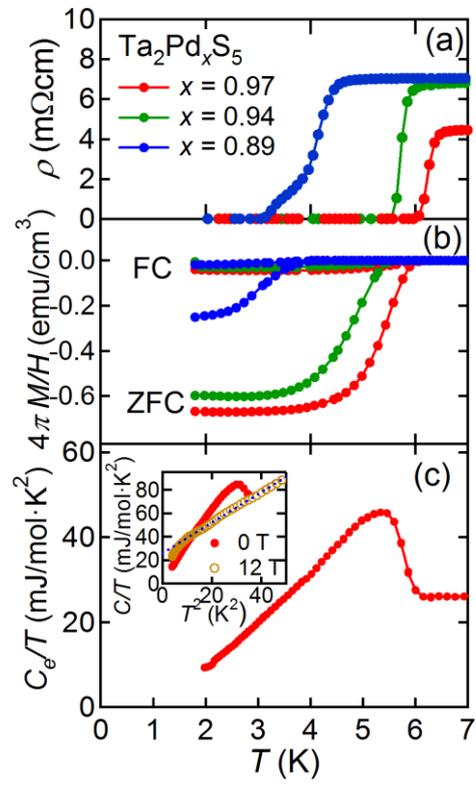

Fig. 2

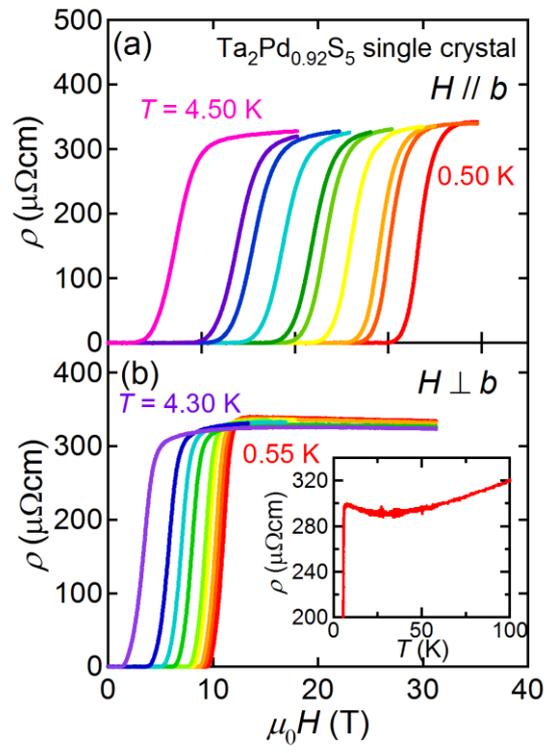

Fig. 3

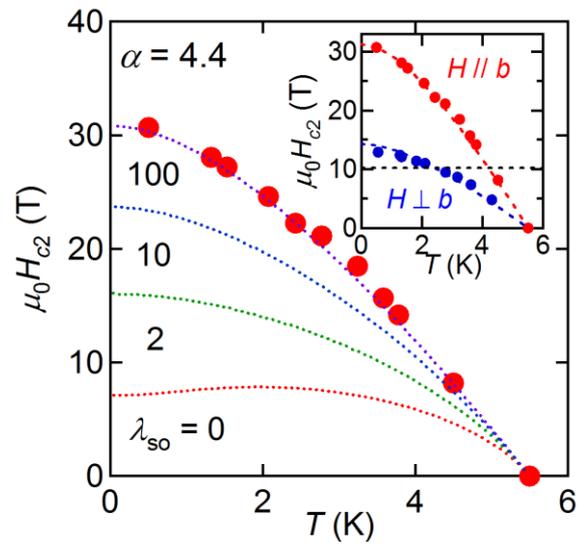

Fig. 4